# Development of a Multiphoton Fluorescence Lifetime Imaging Microscopy (FLIM) system using a Streak Camera

R.V. Krishnan[§] [†], H. Saitoh[★], H. Terada[★], V. E. Centonze[§] and B. Herman[§]

[§] Department of Cellular and Structural Biology, University of Texas Health Science Center, 7703 Floyd Curl Drive, San Antonio, Texas 78229, USA;
[★] Hamamatsu Photonics KK, 325-6 Sunayama-cho, Hamamatsu City, Shizuoka Pref., 430-8587, Japan
[†] Corresponding Author: krsna@uthscsa.edu

*Abstract :*

*We report the development and detailed calibration of a multiphoton fluorescence lifetime imaging system (FLIM) using a streak camera. The present system is versatile with high spatial (~0.2 μm) and temporal (~50 psec) resolution and allows rapid data acquisition and reliable and reproducible lifetime determinations. The system was calibrated with standard fluorescent dyes and the lifetime values obtained were in very good agreement with values reported in literature for these dyes. We also demonstrate the applicability of the system to FLIM studies in cellular specimens including stained pollen grains and fibroblast cells expressing green fluorescent protein. The lifetime values obtained matched well with those reported earlier by other groups for these same specimens. Potential applications of the present system include the measurement of intracellular physiology and Fluorescence Resonance Energy Transfer (FRET) imaging which are discussed in the context of live cell imaging.*

## I.     Introduction

Quantitative live cell imaging is very useful for identifying the various cellular constituents involved in biological phenomena and for probing the molecular interactions among these players. For the study of cell component interactions (i.e. protein-protein interactions), a critical requirement is the ability to visualize living cells with high spatial and temporal resolution.[1,2,3,4] In recent years, for imaging nanometer-scale interactions in cells, the technique of Fluorescence Resonance Energy Transfer (FRET) microscopy has been developed.[5,6,7] The essential requirement for FRET is that there exists a pair of fluorophores (donor and acceptor) between which there is non-radiative transfer of energy when they are in close proximity (~ 1-10 nm). In addition to the separation distance between the donor and acceptor, the efficiency of energy transfer is determined by the angular orientation of the fluorophores in space, overlap between the donor emission and acceptor excitation spectra and the quantum yield of the donor in the absence of acceptor.

Technically, FRET can be monitored by one of the many methods such as sensitized emission, acceptor photobleaching, concentration-dependent depolarization etc. However, since all of these methods rely on intensity-based measurements, they suffer from the need to use optical filters for spectral separation of donor and acceptor excitation and emission and can suffer from problems of spectral cross-talk.[8] This problem, along with the difficulty of regulating the relative concentration of the donor and acceptor fluorophores which in turn affects the efficiency of energy transfer, can complicate the measurement of cell component interactions using FRET. These limitations create the necessity for a non-invasive method that is devoid of such spectroscopic artifacts (as commonly encountered in intensity-based methods) and which can provide an extra dimension of information in cellular imaging. This requirement is fulfilled by Fluorescence Lifetime Imaging Microscopy (FLIM) methods, which measure the intrinsic fluorescence lifetime of the fluorophore in intact cellular environments. When coupled with high temporal and spatial resolution optical microscopy, lifetime imaging provides another dimension of information in the cellular imaging context.[9,10,11,12] This becomes particularly advantageous in situations where the imaged fluorescent proteins are spectrally similar. FLIM methods are insensitive to the spectroscopic details of the fluorophore yet can still provide distinct lifetimes characteristic of the individual fluorophores attached to each protein (e.g. donor and acceptor).

A variety of FLIM methods have been developed in the past decade for measuring various intracellular parameters by monitoring changes in fluorescence excited state lifetime. These have been broadly classified into two major methods: time domain and frequency domain FLIM. The former method relies on measuring the fluorescence intensity decay (~ ns timescales) of the fluorophore when it is excited by a pulsed light source, whereas the latter methods depend on lifetime calculations obtained from the measured changes in amplitude and phase of sinusoidally modulated light before and after passing through the specimen. For a detailed discussion of these methods and a comparative survey of relative merits/demerits of these methods, the reader is referred to earlier review articles.[9,10,11] Despite the availability of current FLIM methods for cell biological applications, a continual requirement that has pervaded this field is to achieve high spatial and temporal resolution in live cell applications. Conventional time-domain FLIM methods such as multigate detection, single photon counting etc., have a typical temporal resolution of a few hundred ps – a limitation imposed primarily by the detectors used in these systems. Some authors improve the accuracy of the data using algorithms after the data collection but this requires certain assumptions about the behavior of the fluorophores inside the cell and may not be accurate.[13] With the advent of multiphoton microscopy, it is now known that combining multiphoton excitation (near infra red: 800-1000 nm) with fluorescence lifetime imaging can yield a high degree of spatio-temporal resolution with some added advantages.[14] These include an increase in spatial resolution as compared to wide-field single photon excitation, deep-tissue imaging, a decrease in photodamage, substantial reduction in spectroscopic cross-talk and better reliability in the information content obtained. Despite these advantages, multiphoton FLIM has not been prominently used in biomedical research as is evidenced by the limited number of literature reports. We present here the development of a new FLIM system that: (i) improves the temporal resolution of the current FLIM systems without sacrificing spatial resolution and without the need of application of post-data acquisition numerical methods, and (ii) minimizes photobleaching of the sample (which improves the reliability of calculated lifetime values) in real-time measurements. The present system combines multiphoton fluorescence imaging and streak-camera based fluorescence lifetime imaging microscopy (Streak-FLIM). Although streak cameras have been used in studies of semiconductor phenomena and picosecond spectroscopy, they have not yet been employed for FLIM in biomedical applications.[15,16,17] In this context, the present multiphoton-Streak-FLIM system is unique and versatile, achieving excellent spatio-temporal resolution and to the best of our knowledge, is the first of its kind developed for biomedical applications.

## II. Principle of Streak Imaging

A fluorescent molecule in an excited state releases its energy as fluorescence emission during its de-excitation to the ground state. This decay of fluorescence intensity of a fluorophore in the excited state can be represented mathematically by a single exponential function:

$$I(t) = I_o \exp(-t/\tau) \qquad (1)$$

where $I(t)$ is the measured intensity, $I_o$ is the intensity at time $t=0$ and $\tau$ is the fluorescence lifetime, characteristic of the molecule. From the above equation, $\tau$ is defined as the time it takes for the initial intensity ($I_o$) to decay to $1/e$ (37%) of its initial value. The above expression has to be modified for multi-exponential fluorescence decays of the fluorophore.

In conventional time-domain FLIM methods such as multi-gate detection, the lifetime is extracted by measuring the fluorescence signal in at least two different time-gated windows. In streak imaging, an optical 2D image with spatial axes (x,y) is converted into a streak image with temporal information and with the axes (x,t). Figures 1 and 2 demonstrate schematically the principle of streak camera image conversion, the general arrangement of the StreakFLIM system (Figure 2A) and the optical layout of the system (Figure 2B). The basic principle of streak imaging can be understood as follows. The streakscope consists of a photocathode surface, a pair of sweep electrodes, a microchannel plate (MCP) to amplify photoelectrons coming off the photocathode and a phosphor screen to detect this amplified output of MCP . Two beam scanner mirrors are employed for scanning along the x and y directions in the effective field of view (~ 40 x 40 µm) (Fig. 2B). The beam scanner G1 scans a single excitation spot along a single line across the x-axis. The width of this line is determined by the photocathode slit at the entrance of the streak camera and the rate of scanning is 1 ms (slow scan mode) in one direction. For 1x1 binning (658x494), this represents a scan of 658 points in a single line. Individual optical pulses (fluorescence emission) are collected from every point along a single line as the beam scanner G1 scans across the x axis. When these optical pulses strike the photocathode surface, photoelectrons are emitted. As the photoelectrons pass between the pair of sweep electrodes a high voltage, synchronized with the excitation pulse, is applied to these sweep electrodes. This sweep voltage steers the electron paths away from the horizontal direction at different angles depending on their arrival time at the electrodes. The photoelectrons then get amplified in MCP and reach the phosphor screen forming an image of optical pulses arranged in vertical direction according to the time of their arrival at the sweep electrodes. The earliest pulse is arranged in the uppermost position and the latest pulse is in the bottom most portion of the phosphor image. The resulting streak image has space as the x-axis and time as the y-axis (Fig.1). When a synchronous y-scanning (G2) is carried out on the region of interest, the above streak imaging process gives a complete stack of (x,t) streak images. This stack contains the complete information of optical intensity as well as the spatial and temporal information from the optical image. Numerical processing of all these streak images ( i.e the exponential decay profiles at every pixel of the raw streak image) gives the final FLIM image. Every pixel in the FLIM image now contains the lifetime information (in contrast to intensity information in optical image). In this way, complete information on fluorescence decays is obtained on a per-pixel basis.

## III.    Experimental Setup

A functional layout of the Multiphoton Streak-FLIM system is given in Figure 2A. A more detailed optical path and data acquisition schematic is presented in Figure 2B.

*A. Pulsed Laser excitation*

The laser system (Model Mira 900, Coherent Inc.) consists of a Titanium: Sapphire gain medium, providing mode-locked ultrafast femtosecond pulses with a fundamental frequency of 76 MHz. The tunability between 700 and 1000 nm allows a wide range of fluorophores to be studied by means of multiphoton excitation. All the measurements reported in this work were carried out at 860 nm. Crucial elements in FLIM measurements are the width of the excitation pulses and the stability of the mode-locked excitation pulse during the course of the experiment. The mode-locked pulses from the laser (width ~ 200 femtoseconds) were found to be very stable during the duration of the measurements. Current limitations of the triggering unit (sweep rates) in the streakscope require operating with repetition rates lower than the fundamental Mira output. The optimum repetition rate of the sweep circuit of the current streak system is 1 MHz. To accomplish the lower repetition rates required for Streak FLIM imaging, we employed a pulsepicker (Model 9200, Coherent Inc.) which uses the input signal from the Mira's fast photodiode output to produce variable repetition rates ranging from 146 kHz to 4.75 MHz. The pulsepicker employs a high-speed acousto-optic modulator (Tellurium Dioxide crystal) to extract a single pulse from the input Mira pulse train. It derives its synchronization signal from the 76 MHz photodiode output of the Mira laser. Besides satisfying the requirements of the triggering unit, the pulsepicker serves another important purpose. By reducing the repetition rate and thereby decreasing the number of excitation pulses, the specimen is not exposed to unwanted excitation pulses and hence is protected from undue photobleaching. As will be seen in the next section, the two photon excitation probability is very well regulated by the pulsepicker. In optimum laser operation conditions, the output pulses of the pulsepicker have a repetition rate of 1 MHz and power typically around 15 mW. Care was taken to ensure precise optical alignment of laser and pulsepicker and of the FLIM optics.

*B. Multiphoton and FLIM optics*

Beam steering optics cause negligible absorption/distortion of the laser beam. The laser output power (~1.75 W at 860 nm) cannot be used as such for biological imaging and has to be decreased substantially for experiments using living cells. It is pertinent to mention here that the pulsepicker is preferred to a simple neutral density filter to attenuate power in the context of multiphoton imaging. This can be understood by considering the following equation for the number of photons absorbed per fluorophore per excitation pulse:

$$N_a \sim [(p_o^2 \cdot \delta)/(\tau_p f_p^2)] \cdot [(NA)^2 / 2\hbar c\lambda]^2 \qquad (2)$$

where $p_o$ is the average laser power, $\delta$ is the fluorophore's two-photon absorption cross-section at wavelength $\lambda$, $\tau_p$ is the pulse duration (~200 fs), $f_p$ is the laser repetition rate, NA is the numerical aperture of the focusing objective, $\hbar$ is the Planck's constant and c is the velocity of light.[4] As can be seen from the above equation, for a constant pulse duration, the two-photon excitation probability increases by either increasing the average laser power or by deceasing the laser repetition rate. A neutral density filter attenuates the laser power regardless of the repetition rate and thereby decreases the two-photon excitation probability according to Equation 2. On the other hand, a pulse picker can increase the two-photon excitation probability by decreasing the laser repetition rate while keeping the average laser power the same. This justifies the use of pulsepicker instead of a neutral density filter in both the multiphoton and FLIM paths in the present system for optimizing laser power at the sample. Regardless of the method adopted to optimize the effective laser power at the sample, it is important to minimize the photobleaching and other photodamage to the sample by laser excitation. It is known that longer wavelength excitation (as in two-photon excitation) considerably reduces photodamage to cells and also cellular autofluorescence since absorption of cellular proteins in this wavelength range is less than that in the near UV region. The operating repetition rates for the multiphoton path can be varied up to 4 MHz depending on the excitation power requirements. In the FLIM optical path, the output of the pulse picker (typically 1 MHz) passes through a beam expander (3X) to achieve a large numerical aperture at the back focal plane of the microscope objective.

The entire optical alignment of the FLIM elements is done with respect to the optical axis of microscope objective (Figure 2B). The horizontal galvo mirror (G1) is the beam scanner used to scan the laser spot in the horizontal direction (x-axis) and is enabled by a triangular wave from a pulse generator (Hewlett Packard 8112A). The scanning length can be varied by varying the peak-to-peak voltage amplitude of the triangular wave. A raster scanning mechanism is employed for both the x and y axes (G1 and G2) and their linearity performance characteristics will be discussed in later sections. A focusing relay lens (L1) positioned next to G1 facilitates sharp focus of this scanned beam onto the dichroic D1. Dichroic D1 is a short-pass filter with a cut-off at 750 nm. Light reflected from D1 falls on the vertical galvo mirror (G2) which is responsible for y-scanning in the streak image. The fluorescence emission from the microscope is collected by G2 and transmitted through D1 for detection by the streakscope (Figure 2B). The imaging lens prior to the streakscope photocathode critically influences collection efficiency. It is therefore very important to precisely align the imaging lens with bright fluorescent specimens as part of system calibration.

*C. Microscope*

The left port of an Olympus IX70 microscope was assigned for the multiphoton path while the Streak FLIM path was assigned the right port of the microscope (Figure 2A). Conventional fluorescence microscopy can be performed with an arc lamp at the rear port. Laser scanning/descanning is effected in a FluoView (Olympus) scan module modified for multiphoton imaging. A specially designed 63X (1.2 N.A, IR) water immersion objective was used for all the measurements reported in this paper. As can be seen from Equation 2, the high numerical aperture provides for spatial confinement of the excitation power in a small focal volume thereby increasing the two-photon excitation probability. Experiments were done with cover glasses (thickness of 0.17 mm) and specimens were either solutions sandwiched between two cover slips or fixed cellular specimens mounted on standard microscope slides.

*D.     Detection System*

A standard head-on photomultiplier tube was used for detection of fluorescence emission in the multi-photon path and a microchannel plate was used in the Streak-FLIM path. The Streakscope (C4334,Hamamatsu,) used in the present system has a temporal resolution ~50 ps and has a photocathode dark current three orders of magnitude smaller than that in the photomultiplier tubes used in SPC systems, thus offering a high signal-to-noise ratio (SNR) for measuring even weak fluorescence signals. The electrical output of MCP is converted to an optical output (streak image) on the phosphor screen. This image is then read out by a fast CCD camera (ARGUS/HiSCA,Hamamatsu,) which is fiber-optically coupled to the streakscope and then undergoes analog-to-digital conversion. An important feature of this camera is that it employs a CCD which offers exceptionally high-speed and high sensitivity detection. A maximum sampling rate of ~ 500 frames/sec can be achieved with this camera for single wavelength fluorescence measurements which is more than ten times higher than the normal video rate. This feature makes the present system unique for rapid data acquisition with a better

SNR than expected from conventional CCD cameras. Faster frame acquisition can be achieved by binning the pixels at the expense of loss of spatial/temporal resolution; thus a balance between the speed of the data acquisition and required resolution has to be determined in every measurement. This relationship can also vary between samples depending on the strength of their fluorescence signals. The camera's maximum number of effective pixels is 658 x 494 in the unbinned condition; a maximum of 32 x 32 binning (in both space and time axes of the streak image) can be achieved. The system can be operated in two scanning modes: slow scan with a better linearity (12 bits) or fast scan (10bits). The multi-photon data acquisition is automated through FluoView (Olympus) software. The FLIM data acquisition is automated through AquaCosmos software (Hamamatsu). This involves numerical processing of the raw streak images on a per pixel basis and conversion into a FLIM image corresponding to the optical image of the specimen. The factors affecting the FLIM calculations will be discussed in the next section.

## IV.     Instrument Performance

Primary requirements from a robust lifetime measurement system are: accuracy, wide dynamic range, high sensitivity and reproducibility. We carried out many measurements in the Streak-FLIM system to validate the above requirements.

### A.     *System calibration*

Standard solutions of various fluorescent dyes were used for calibration of the system and for optimizing relevant parameters for a typical measurement. Rhodamine 6G and Rose Bengal solutions were prepared in different solvents as detailed in Table 1. These dyes are reported to have mono-exponential fluorescent decays and therefore display a single lifetime.[18] The choice of standard dyes we employed also allowed calibration of the system for measurement of lifetime values ranging from a few hundred picoseconds to a few nanoseconds. Figure 3 illustrates the lifetime histograms for the three standard fluorophore dyes used. Table1 also provides the error analysis for each histogram and compares the calculated lifetime values obtained with the Streak-FLIM system with those reported in the literature. There is a very good agreement between the calculated lifetimes with those in the literature, demonstrating the accuracy of lifetime determination of the Streak-FLIM system.

Accuracy of the lifetime measurements is largely governed by the uniformity of FLIM images. Lifetime images of the standard solutions that we employed exhibited a high degree of uniformity and thus provided high accuracy in the calculations of lifetime. To further characterize the system and to examine the relationship between the spatial and temporal resolution in measurements involving specimens with defined structure, we used a constellation microsphere mixture (Molecular Probes Inc., Catalog # C-14837). Figure 4 illustrates a representative multiphoton and FLIM image for green and red microspheres, and their mean lifetime values. Figure 5 demonstrates FLIM images for the green microsphereat different exposure times (i.e., proportional to the rate of data acquisition). The duration of a single (x,t) scan is typically 2 ms; to achieve good SNR, this signal can be integrated over a defined time period determined by the user. For scanning an area of ~ 40 μm (~ 300 slit lines scanned in the y direction) from a reasonably bright specimen (giving a signal well below the saturation threshold of the MCP output), it is therefore possible to achieve an average data acquisition time of less than a second when employing 32x32 binning. This rapid data acquisition capability provides great advantage in imaging biological specimens where prolonged exposure to incident light usually causes irreversible photobleaching, phototoxicity and photodamage. However, as long as the specimen does not suffer from photobleaching, it is always preferable to integrate the signal over a longer period of time to achieve good SNR. This fact is illustrated in Figure 5, which demonstrates lifetime histograms for the green microsphere measured using different integration times for the (x,t) scan. With this sample, the lifetime values are nearly constant for exposure times varying from 36 ms to 210 ms, demonstrating rapid data acquisition capabilities of the Streak FLIM system.

Average intensity per streak frame (AISF) is the key parameter that determines the signal-to-noise ratio in the Streak FLIM system and this parameter has to be optimized in every measurement to get reproducible lifetime values. Input laser power, exposure time per streak frame, beam scanning speed and MCP gain are the critical parameters that can influence AISF. Figure 6 demonstrates the variation in AISF as a function of exposure time and MCP gain (voltage on the MCP). A nonlinear response of AISF as a function of MCP gain is typical of photomultiplier devices. However, we observe that the AISF scales linearly with exposure time in the slow scan mode. Since the slope of the triangular waveform (which excites the beam scanner G1) determines the rate of scanning (2 ms in our case), the slight nonlinearity observed in fast scan mode is due to the nonlinearity in the analog derivative of this waveform specific to fast scan mode. The exposure times and MCP gain values employed are representative of typical imaging conditions used for experiments.

One potential artifact that can affect the accuracy of lifetime determination in a streak image is the presence of a large number of saturating pixels, which would lead to the calculated lifetimes being higher than the real lifetime values. Care was taken to analyze the exponential decays in different frames (i.e. x-y scans) and the images with exponential decay artifacts (those shifted to right because of saturated pixels) were discarded. Another way to increase AISF is to perform binning on the camera itself although this will be at the expense of spatial/temporal resolution. By binning only along the time-axis, a reasonable compromise can be achieved between spatial and temporal resolution.

The raw streak images were numerically processed by AquaCosmos (Hamamatsu) software and the lifetime calculations were done in one of two modes: linear interpolation (low precision mode) or exponential interpolation (higher precision mode). In the former approach, the lifetime value is taken to be the difference between two time points, the maximum intensity ($I_o$) and 37% (1/e) of the maximum intensity. In the case of exponential interpolation, the time profile data from the streak image (i.e vertical line along the time axis) is fitted to an exponential decay equation to obtain the lifetime value. For the simple case of mono-exponential decay, Equation 1 holds. In practice, the calculated mean lifetime values were very similar ($\tau_{lin}$ vs. $\tau_{exp}$) when either of these approaches was used, although the standard deviation of the lifetime values was smaller using exponential interpolation. It is relevant here to emphasize that for both of the above approaches, the confidence level in the obtained lifetime values is largely governed by the uniformity in the FLIM image. Besides uniform light illumination and sample preparation conditions, FLIM image uniformity also depends on the optimization of a variety of factors as was discussed previously. Since the streak cameras are characterized by a high temporal resolution (~ 50 ps) and a high dynamic range (10000: 1), even weak signals can be detected easily.[19]

## B. Biological applications

To demonstrate the applicability of the Streak-FLIM approach to cellular specimens, we chose to image a slide of mixed pollen grains. As can be seen from Figure 7, lifetime values as small as 0.20 –0.34 ns can be measured with good spatial resolution ( ~0.2 µm). In Table 1, we report mean lifetime values; it should be noted that these samples have been reported to have multi-exponential decays.[20] Figure 8 shows the multiphoton image, FLIM image and lifetime histogram of a fixed specimen of baby hamster kidney (BHK) cell expressing mitochondrially-targeted enhanced cyan fluorescent protein (mECFP). As it can be seen from the figure, the lifetime image displays good spatial and temporal resolution and the accompanying lifetime histogram shows a relatively narrow distribution of lifetimes around the mean value (~2.8 ns). This value agrees well with values obtained earlier for the same probe.[21] It has been previously observed that lifetime measurements of ECFP yielded a two component-lifetime.[21] Exponential interpolation did demonstrate two lifetimes ( ~2.9 and 2.2 ns); however, here, we assume a single exponential decay for ECFP in our analysis and figure 8 shows the mean lifetime histogram of single component fitting. We also measured the mean lifetime of enhanced green fluorescent protein (not shown) expressed in the cytosol of BHK cells. This value was also in very good agreement with literature reports of lifetime values of 2.8-3.0 ns for the same probe.[22] As the fluorescent proteins are gaining reputation as non-invasive intracellular probes for a variety of biological phenomena, the fluorescence lifetime measurements of these probes in different cellular environments will be vital for understanding both the photophysical properties of the probe as well as the biological phenomena in living cells.

In conclusion, we report here the development and detailed calibration of a multiphoton fluorescence lifetime imaging system using a streak camera. The present system is versatile with high spatial and temporal resolution allowing rapid data acquisition as well as reliable and reproducible lifetime determinations. We demonstrate the applicability of the system to cellular specimens and the lifetime values we obtain agree very well with those reported in the literature. The system holds promise as a reliable tool to investigate a number of biological phenomena more accurately than the conventional intensity-based microscopy methods as well as the other FLIM methods. Future improvements to the system will include the ability to analyze multi-component lifetimes to address more complex situations where the fluorescence decays are not single exponentials. These analyses will shed more light on the subtle photophysical schemes of fluorophores as well as on the dynamics of protein/cell component interactions within living cells.

**Acknowledgements**

We thank Atsushi Masuda for preparing BHK cell specimens and Motoyuki Watanabe for useful discussions on FLIM calculations.

_________________________________

**Figure Captions**

**Figure 1:** *A schematic of the principle of streak imaging.* An optical image (A) with intensity information at every pixel is converted to a streak image (B) with spatial information as the horizontal axis and time as the vertical axis. Every point in B provides three-dimensional information- intensity, space and time, corresponding to every pixel in A. **S** : Slit; **L** : Imaging lens; **PC**: Photocathode; **SE**: Sweep electrodes; **M** : Microchannel Plate; **PS**: Phosphor screen. In this schematic, three optical pulses arrive at the slit with varying intensity and which vary slightly in terms of time and space. As the corresponding photoelectrons from PC pass between the pair of sweep electrodes, the applied sweep voltage steers the electron paths away from the horizontal direction at different angles depending on their arrival time at the electrodes. The amplified electrons reach the PS forming an image of three optical pulses arranged in vertical direction according to the time of their arrival at the sweep electrodes. The earliest pulse is arranged in the uppermost position and the latest pulse is in the bottom most portion of the phosphor image. The resulting streak image (B) has space as the x-axis and time as the y-axis.

**Figure 2:** *Functional (A) and Optical (B) layout of the Multiphoton StreakFLIM system.* G1: Horizontal Galvo Mirror; G2 : Vertical Galvo Mirror ; L1: Focusing relay lens; C: Coupling lens for sideport optics; I : Imaging lens; O : Microscope Objective lens; D1 : Dichroic; PC :Photocathode;

**Figure 3:** *Lifetime images & histograms for standard fluorophore solutions.* (a&b) Rhodamine 6G in ethanol (c&d) Rose bengal in acetone (e&f) Rose bengal in ethanol. Solid lines are Gaussian fits to the data. Every pixel in the lifetime images (a,c and e) gives the lifetime of the fluorophore at that particular position in space (x,y).

**Figure 4:** *Multiphoton (a, d), and Streak-FLIM (b ,e) images and Lifetime histograms (c, f) for green and red latex microsphere fluorescent microspheres. The FLIM images were obtained in the unbinned condition. The apparent difference in multiphoton and FLIM images is due to the difference in effective area in the respective detectors and hence in the pixel sizes of their fields of view. Mean values of lifetime for the green and red microsphere are 1.50 ns and 2.13 ns respectively.*

**Figure 5:** *Rapid data acquisition and temporal resolution of the StreakFLIM system. The measurements were done on a green microsphere with 300 streak frames for the entire microsphere. Lifetime histograms shown correspond to different data acquisition times/streak frame namely (a) 210 ms; (b) 146 ms; (c) 73 ms; and (d)36 ms. The calculated lifetimes in all the cases were in the range of 1.50 ns ± 0.15 ns irrespective of the data acquisition times, thereby showing the stability of the system even at fast acquisition speeds. It can be observed that the width of distribution becomes narrower with an increase in data acquisition time from 0.30 ns (36 ms) to 0.22 ns (210 ms).*

**Figure 6:** *Variation of average intensity per streak frame in the slow vs. fast scan mode. (a) Normalized average intensity varies linearly with the exposure time/streak frame although the fast scan shows a slight nonlinearity for higher exposure times ( see text). (b) The normalized average intensity varies nonlinearly with the MCP gain which reflects the typical nonlinear response of photomultiplier tube devices.*

**Figure 7:** *Multiphoton (a,b) and StreakFLIM (c, d) images of two different sections of mixed pollen grain. The calculated lifetime value inside the pollen grain section is 0.36 ± 0.03 ns (e). The standard error in the lifetime calculation is less than the optimum time resolution of the streak camera (0.05 ns).*

**Figure 8:** *Multiphoton image(a), StreakFLIM image (b) and lifetime histogram (c) of a Baby Hamster kidney (BHK) cell transfected with mitochondria-targeted enhanced cyan fluorescent proteins (mECFP). The calculated lifetime value inside the mitochondria is 2.71 ± 0.25 ns.*

**Table 1 Legend**

The standard solutions were prepared by dissolving the dye powders in different solvents and diluted stocks were prepared for calibration. Calculated lifetime values did not change with the concentration of the dye in the nominal range of a few hundred nM to a few hundred μM. All the measurements were carried out at 200 μM concentration for system calibration. Raw streak images were obtained and the mean lifetimes were calculated by the methods described in the text. Monoexponential decays were assumed for all the calibration samples as reported in literature and the values reported here are all mean lifetime values. Standard deviations were obtained by fitting the lifetime histograms to normal distribution.

**Table 1**: Measured Lifetimes in the Streak-FLIM system

| Samples | | **Measured Lifetimes** | | **Literature** | |
|---|---|---|---|---|---|
| | | $\tau$ (ns) | *Standard deviation* (ns) | $\tau$ (ns) | *Reference* |
| Calibration Probes | Rhodamine 6G / Ethanol | 2.97 | 0.14 | 3.0 | 17 |
| | Rose Bengal / Acetone | 2.22 | 0.13 | 2.4 | 17 |
| | Rose Bengal / Ethanol | 0.74 | 0.04 | 0.8 | 17 |
| Latex Microspheres | Green | 1.49 | 0.15 | ___ | ____ |
| | Red | 2.01 | 0.26 | ___ | ____ |
| Cellular specimens | Pollen grain | 0.35 | 0.08 | 0.5 | 19 |
| | mECFP | 2.71 | 0.25 | 2.68 | 20 |
| | EGFP | 3.02 | 0.22 | 2.8 – 3.0 | 21 |

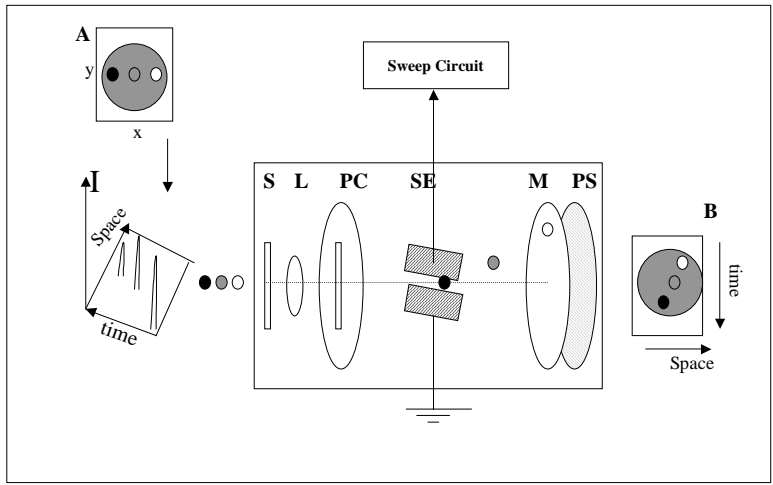

Fig.1

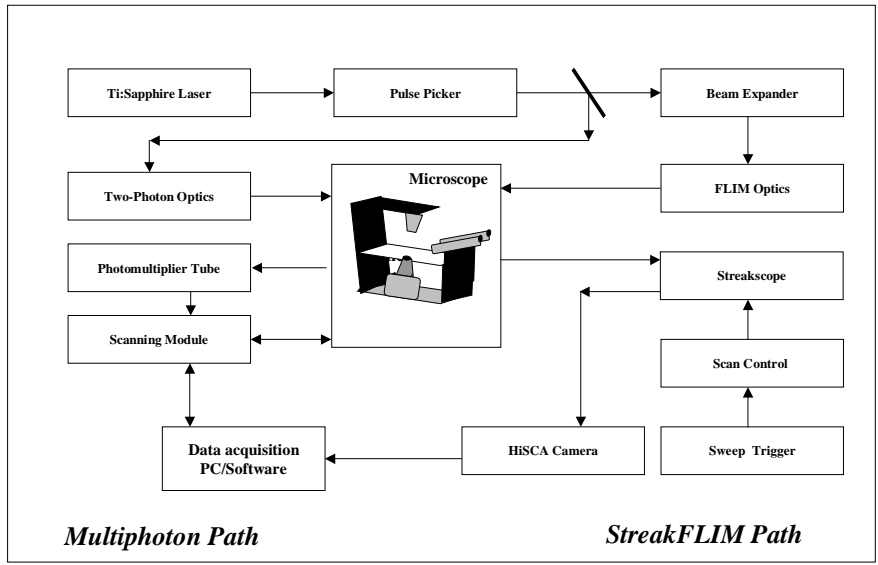

Fig.2A

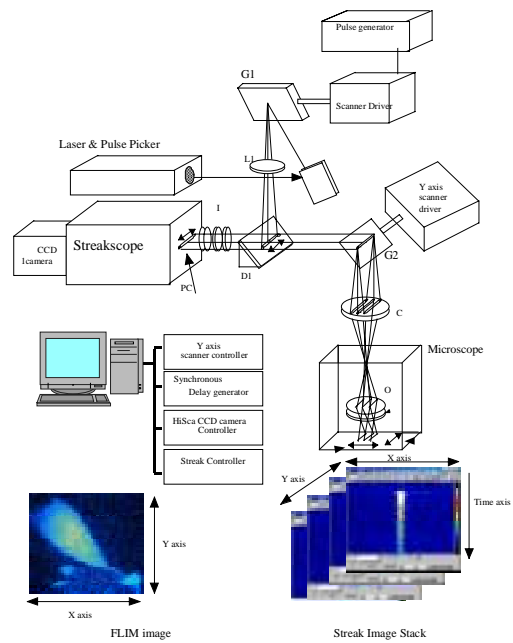

Fig.2B

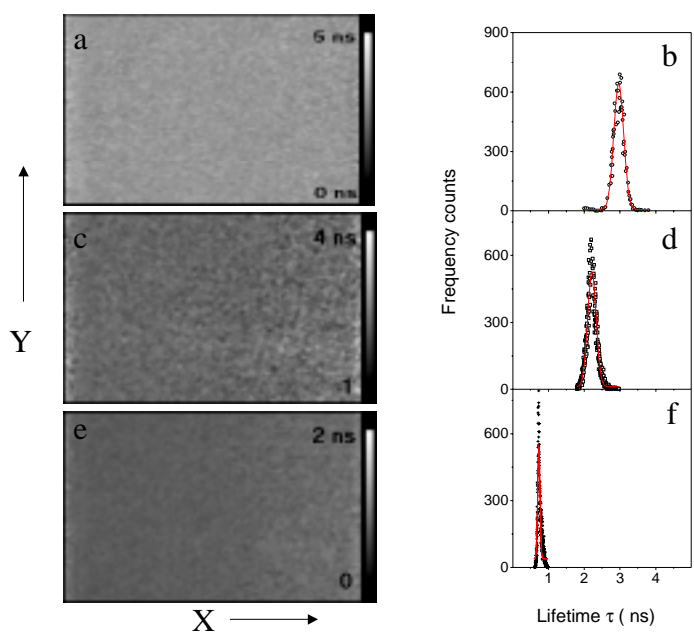

Fig.3

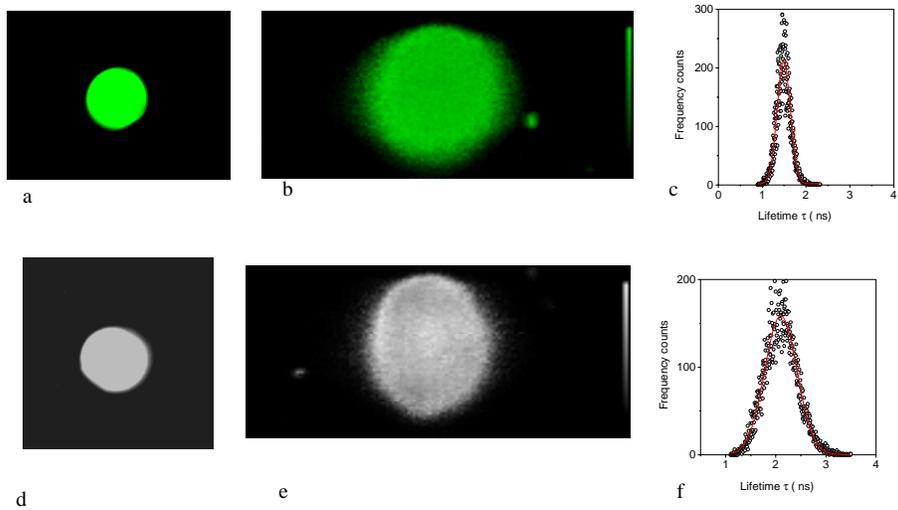

Fig.4

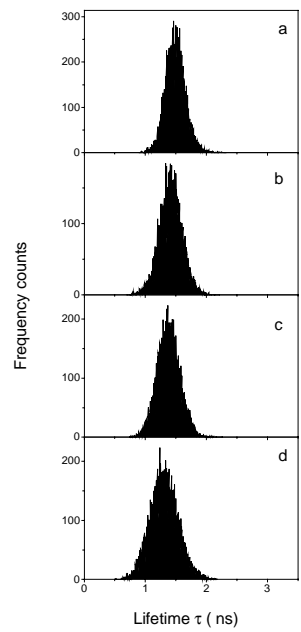

Fig.5

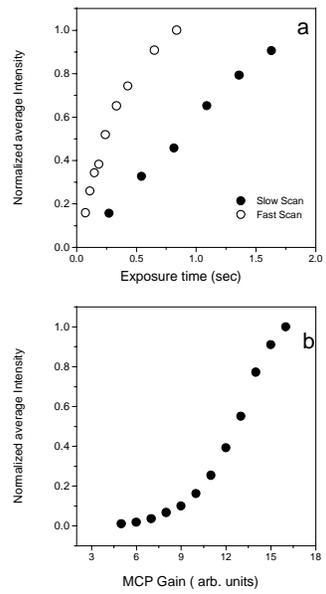

Fig.6

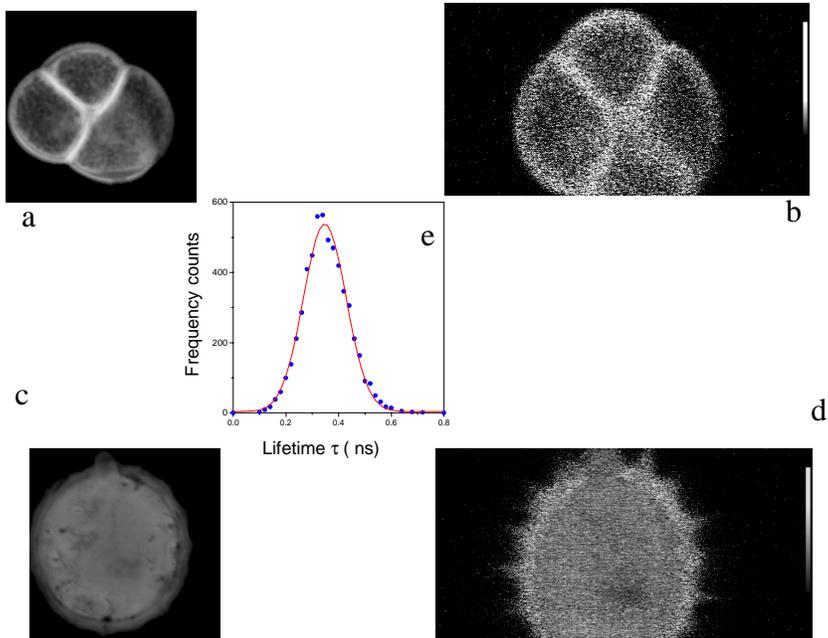

Fig.7

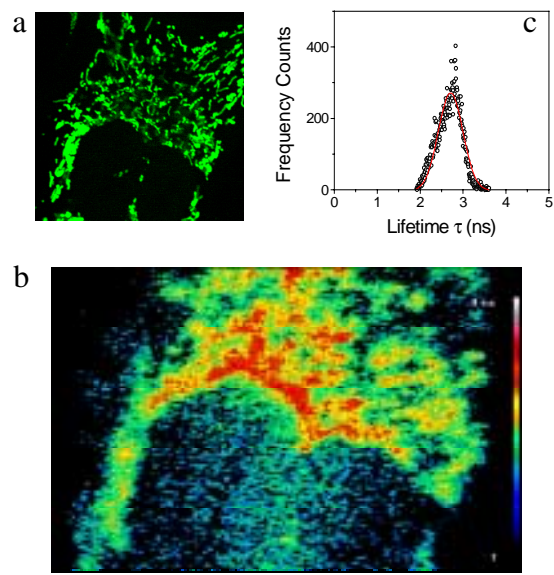

Fig.8